\documentclass{article}    

\usepackage{graphicx}          

\usepackage[latin1]{inputenc} 
\usepackage{amssymb}          
\usepackage{amsmath}          
\usepackage{amsfonts}

\newtheorem{thmsec}{Theorem}[section] 		
\newtheorem{prosec}{Proposition}[section] 	
\newtheorem{lemsec}{Lemma}[section] 		
\newtheorem{corsec}{Corollary}[section] 	

\newtheorem{dfnsec}{Definition}[section]

\newcommand{\ket}[1]{|#1\rangle}
\newcommand{\bra}[1]{\langle #1|}

\newcommand{\Hi}{\mathcal{H}}

\newcommand{\C}{\mathbb{C}}

\newcommand{\Ec}{\mathcal{E}}

\newcommand{\tr}{\textrm{tr}}

\def\smallfrac#1#2{{\textstyle\frac{#1}{#2}}}

\newcommand{\qed}{\hfill $\Box$ \vskip 2ex}

\newcommand{\beq}{\begin{equation}}
\newcommand{\eeq}{\end{equation}}
\newcommand{\beqa}{\begin{eqnarray}}
\newcommand{\eeqa}{\end{eqnarray}}
\newcommand{\beqan}{\begin{eqnarray*}}
\newcommand{\eeqan}{\end{eqnarray*}}
\newcommand{\bea}[1]{\begin{eqnarray}\label{#1}}
\newcommand{\eea}{\end{eqnarray}}

\newcommand{\be}[1]{\begin{equation}\label{#1}}
\newcommand{\ee}{\end{equation}}
\newcommand{\bp}[1]{\begin{prosec}\label{#1}}
\newcommand{\ep}{\end{prosec}}
\newcommand{\epr}{\qed}
\newcommand{\bpr}{\noindent{\em Proof: }}

\newcommand{\DD}{\mathfrak{D}(\Hi)}
\newcommand{\HH}{\mathfrak{H}(\Hi)}
\newcommand{\PP}{\mathfrak{P}(\Hi)}
\newcommand{\UU}{\mathfrak{U}(\Hi)}
\newcommand{\BB}{\mathfrak{M}(\Hi)}
\newcommand{\reach}{\mathfrak{R}}
\renewcommand{\P}{\mathbb{P}}

\newcommand\M{ M_{k}^{\phantom{\dagger}}}				
\newcommand\Ma{	M_{k}^{\dagger}}						

\renewcommand\cal{\mathcal}

\begin{document}

\title{Discrete-Time Controllability\\ for Feedback Quantum Dynamics} 

\author{Francesca Albertini\thanks{Dipartimento di Matematica Pura ed Applicata, Universit\`a di Padova, via Trieste 63, 35131 Padova, Italy, {\tt albertin@math.unipd.it}} { }%
and    
Francesco Ticozzi\thanks{Dipartimento di Ingegneria dell'Informazione, Universit\`a di Padova, via Gradenigo 6/B, 35131 Padova, Italy, {\tt ticozzi@dei.unipd.it}}}               
\maketitle

\begin{abstract} 
Controllability properties for discrete-time, Markovian quantum  dynamics are investigated. We find that, while in general the controlled system is not finite-time controllable, feedback control allows for arbitrary asymptotic state-to-state transitions. Under further assumption on the form of the measurement, we show that finite-time controllability can be achieved in a time that scales linearly with the dimension of the system, and we provide an iterative procedure to design the unitary control actions. 
\end{abstract}


\section{Introduction}

For any controlled system, an in-depth study of its controllability properties under the available control capabilities is the necessary premise to the  design of effective controls addressing some given task. For quantum systems, in particular, controllability properties have been studied mostly considering continuous-time models in the presence of open-loop, coherent controls \cite{dalessandro-book,albertini-notions,schirmer-controllability,altafini-controllability,altafini-markovian,altafini-open,thomas-controllability}. In this setting, the evolution is deterministic and the problem can be studied with the tools of geometric control theory. 

Indeed, for classical deterministic systems it makes little sense to distinguish open-loop and feedback controllability: the fact that the control law can benefit from partial or complete information on the system trajectory does not modify the reachable set from a given initial state. In the quantum case, however, the introduction of measurements alone modifies the dynamical model by introducing a stochastic behavior, which has to be carefully taken into account. Considering the ``open-loop'' effect of measurements is not enough: the ability of {\em conditioning the control choice on the measurement outcomes changes significantly the controllability properties}, and in particular the set of reachable density operators as it will be argued later.  

Continuous-time controllability of {\em open-loop} quantum dynamical semigroups have been studied in \cite{altafini-markovian,altafini-open,thomas-controllability}. Some preliminary ideas about discrete-time, open-system controllability have been also previously explored in \cite{kraus-controllability}. In that case, however, no reference to a specific set of control capabilities has been made (open-loop,closed-loop, coherent, incoherent, measurement-based control,...), the main focus being on the existence of general open-system dynamics connecting any given pair of states.

In this paper we investigate the controllability properties of controlled, Markovian {\em discrete-time quantum dynamics} in open and closed loop. As a preliminary step, we will argue that a discrete-time system obtained by sampling inherits the open-loop controllablity properties from the continuous time underlying mode by resorting to previous results by Sontag \cite{sontag-1,sontag-2}. Open-loop controllability is a generic property for closed quantum systems, and this motivates our assumption of unitary controllability of the discrete-time systems we consider next. On the other hand, by introducing generalized measurements and closing the loop with conditional control actions, the dynamics drastically changes and our main results shall focus on this setting.
We will present three simple examples illustrating how: (i) conditioning the control action on the outcome of a measurement influences the reachable sets of a controlled open-system evolution; however, in general (ii) feedback control does not in general ensure finite-time, state-to-state controllability; and (iii) feedback control does not allow for engineering of arbitrary dynamics.
Next, we will prove that, under generic condition on the chosen measurement, feedback allows for {\em asymptotic state-to-state controllability}. Lastly, we will study a particular, yet not so restrictive in practice, class of controlled dynamics that exhibit {\em finite-time feedback state-to-state controllability}. As a byproduct of the proof of finite-time controllability, an explicit way to construct the sequence of control actions is provided. Remarkably,  the (maximum) number of feedback steps needed to obtain any desired state-to-state transition scales linearly with the dimension, namely it is twice the size of the system's Hilbert space.

The paper is structured as follows: after recalling the essential features of quantum systems in Section \ref{prelim} and the relevant notions of controllability in Section \ref{notions}, in Section \ref{openloop} we argue that samples dynamics inherits open-loop controllability from the underlying continuous-time model. Beside being of interest by itself, the ability of enacting arbitrary control actions in finite-time is also a key assumption in Section \ref{sec:feedback}, where we establish under which conditions feedback control ensures asymptotic state-to-state controllability. After presenting the general results on feedback approximate controllability in Section \ref{sec:feedback}, Section \ref{k2} will describe a particular class of dynamics, proving that in this case finite-time state-to-state controllability can be achieved.
 
\section{Discrete-time Quantum Dynamics and Controllability Notions}\label{prelimi}

\subsection{Quantum Systems}\label{prelim}

In this paper we will consider finite-dimensional quantum systems. Let us introduce some basic notation: to the quantum system of interest is associated an Hilbert space $\Hi\sim \C^N.$ In Dirac's notation (see e.g. \cite{sakurai}), vectors in $\Hi$ are denoted by {\em kets}, $\ket{\psi}\in\Hi$, while the linear functionals on $\Hi$ live on the dual space $\Hi^\dag$ and are denoted by {\em bra}, $\bra{\psi}.$ Inner products are then represented by $\bra{\phi}\psi\rangle$ ({\em bra(c)ket}s). $\BB$ denotes the set of linear operators on $\Hi$. Consider $X\in \BB$: its action on bras and kets is defined by $X\ket{\psi}:=X(\ket{\psi}),$  $\bra{\psi}X:=X^\dag(\bra{\psi}),$ where $\dag$ denotes the adjoint of operators (and consistently the transpose-conjugate for their matrix representations). 

Self-adjoint (Hermitian) operators are denoted by $X^\dag=X\in\HH,$ and are associated to {\em observable} variables for the system.
In a quantum statistical framework, a {\em state} for the system is associated to trace-one, self-adjoint and positive-semidefinite operator $\rho.$ Let us denote by $\DD$ the set of states or {\em density operators.}  The subset $\PP \subset \DD$ denotes the set of rank-one orthogonal projectors, the {\em pure states}. $\DD$ is a convex set, whose extreme points are the pure states $\PP,$ and its border $\delta\DD$ contains all the states that are not full rank.

In this paper we will consider {\em generalized measurements}, with a finite number of possible outcomes labeled by an index $k$. Assume a system is in the state $\rho.$ A generalized measurement is associated to a decomposition of the identity $\sum_k \Ma \M = I,\,M_k\in\BB$ that allows to compute the probability of measuring the $k$-th outcome as:
\[\P_\rho(k)=\tr(M_k\rho M_k^\dag),\]
and the conditioned state after the measurement as:
\be{conditional}\rho|_k=\frac{M_k\rho(t)M_k^\dag }{\tr( M_k\rho(t)M_k^\dag)}.\ee
A particular case is represented by direct measurements of observables, or {\em projective} measurements: consider an observable $X\in\HH,$ with spectral representation $X=\sum_k x_k\Pi_k,$ $\sum_k\Pi_k=\sum_k\Pi_k^2=I.$ The eigenvalues correspond to the possible outcomes of the measurement, labeled by $k$, and the probabilities and conditioned states can be computed by the formulas above with $M_k=\Pi_k.$

We shall consider dynamics in the so-called Schr\"odinger picture, where the state is evolving while the observables are time-invariant.
It follows from Schr\"odinger's equation (see next section, equation \eqref{schrod}) that an {\em isolated, closed} quantum system evolves unitarily: in discrete time, this means that for a  sequence of times $t=0,1,2,\ldots,$ with time-intervals normalized to one, we have
\be{unitaria}
 \rho(t+1)=U(t)\rho(t)U^\dag(t), \ee
with $U(t)\in\UU$ for all $t$'s (here $\UU\subset \BB$ denotes the subset of unitary operators). 
In the open quantum system setting, general physically admissible evolutions  are described by linear, Completely Positive and Trace Preserving (CPTP) maps \cite{nielsen-chuang,petruccione}. Any CPTP map $\mathcal T$ via the Kraus-Stinespring theorem \cite{kraus} admits explicit representations of the form 
\begin{equation}
\mathcal{T}[\rho] = \sum_k M_k \rho M_k^\dag \label{eq:KrausMap}
\end{equation}
also known as Operator-Sum Representation (OSR) of $\mathcal T$, where $\rho$ is a density operator and $\{M_k\}$ a family of operators such that the completeness relation
\begin{equation}
\sum_k \Ma \M = I \label{eq:KrausMapConditions}
\end{equation}
is satisfied. We refer the reader to e.g. \cite{alicki-lendi,nielsen-chuang,petruccione} for a detailed discussions of the properties of quantum operations and the physical meaning of the complete-positivity property.
We recall that maps in the form \eqref{eq:KrausMap} preserving the identity, ${\cal T}(I)=I$, are called {\em unital}. It is well known that the OSR of a given CPTP map is not unique: in fact the following holds (see \cite{nielsen-chuang}, {Theorem 8.2}):
\begin{thmsec}{\em (Unitary freedom in the OSR)} 
Assume $\{M_k\}_{k=1}^{m}$ and $\{N_k\}_{k=1}^n$ be OSRs of quantum operations ${\cal E}$ and ${\cal F},$ respectively. If $m\neq n,$ append zero operators to the shortest list so that $m=n.$  Then ${\cal E}={\cal F}$ if and only if there exist a unitary $n\times n$ matrix $U=[u_{k\ell}]$ such that: \[M_k=\sum_{\ell}u_{k\ell}N_\ell.\]
\end{thmsec}
\vspace{-5mm} In the rest of the paper, however, open-system dynamics will be obtained as averages over states conditioned on a {\em given measurement}, followed by unitary control. By averaging over the possible outcomes of a generalized measurement we get:
\[\bar\rho={\cal M}(\rho)=\sum_k\P_\rho(k)\rho|_k=\sum_kM_k\rho M_k^\dag, \]
which is a CPTP map, and physically represents the expected effect of a measurement on the state, when the outcome is not known. The fact  that $\bar\rho\neq\rho$ is a remarkable difference with respect to classical probability. Also notice how, of all the possible OSR associated to the {\em unconditional} ${\cal M},$ only one, $\{M_k\},$ will corresponds to the correct {\em conditional} states via \eqref{conditional}. This means that when considering feedback protocols based on the conditional states (as we do in Section \ref{sec:feedback}), different OSRs are not equivalent, and we have to consider the  fixed OSR associated to the underlying measurement.

\subsection{Notions of Controllability}\label{notions}

When dealing with dynamics depending on external controls, it is of physical interest to know whether or not these controls can be chosen so as to drive the {\em{state}} of our model between two given configurations, either exactly or approximately. Different notions of controllability can be given depending on which is the relevant {\em{state}} for the dynamics. As an example, when dealing with multilevel quantum mechanical systems evolving in continuous time we may look at the evolutions on the complex unitary sphere, on the unitary operations, or on the density matrix operator. More precisely, 
denoting by $H(\vec{u}(t))=H_0+\sum_{j=1}^m H_ju_j(t)$ the Hamiltonian including the controls, and 
considering the system isolated, we can study controllability of the Schr\"odinger equation,
\be{schrod}
 \ket{\dot{\psi}(t)}= -i H(\vec{u}(t))\ket{\psi(t)},\   \  \  \text{ with } \ket{\psi(t)}\in 
  S^{N-1}_{\C}, 
 \ee
  describing the evolution on the complex unitary sphere associated to pure states, or the corresponding equation acting on the propagator,
\[
\dot{X}(t)=-i H(\vec{u}(t))X(t), \   \  \  \text{ with }  X(t)\in \UU,
\]
or again the Landau-von Neumann equation  
\beq\label{lvn} \dot\rho(t)=-i[H(\vec{u}(t)),\rho(t)],\eeq
for the evolution on the density operators.
Thus, according to the problem we are looking at, we may be interested to the action of the same Hamiltonian to either $\ket{\psi},\,X$ or $\rho$. Of course, the controllability properties are connected: notice that if $X(t)$ denotes the solution of the second equation with initial condition $X(0)=I$, then we have $\ket{\psi(t)}= X(t) \ket{\psi(0)}$ and $\rho(t)=X(t) \rho(0) X(t)^{\dag}$. These relations provide some correlations among the different types of controllability. 

In this paper we will deal instead with controlled open quantum models evolving in discrete time on the set of density operators. The dynamics will be generically described by:
\be{generica}
\rho(t+1)= \Ec(\rho(t),\vec{u}(t)),
\ee
with $\rho(\cdot)\in \DD $ and $\vec{u}(t)\in U$, and  where $U$ is the set of controls.
Later we will be precise about the set of controls $U$ and the form of the map $\Ec$. In particular, we will deal with the case where $\Ec$ comes from sampling a continuos time model evolving according
the Landau-von Neumann equation (see (\ref{sampling})), and with the case where $\Ec$ is a CPTP map emerging from measurement and feedback unitary control
(see (\ref{feedback})). Since the subset  $\PP \subset \DD$ of the pure states have a special physical meaning, we introduce the following different definitions of  controllability properties. 

{\em Pure state to Pure state Controllable (PPC) in $T$ steps:} if for every pure initial state $\rho_0=\ket{\psi}\bra{\psi}\in\PP$ there exist a sequence of controlled dynamical maps $\Ec_1,\ldots,\Ec_T\in{\cal C}$ such that any other pure $\rho_f\in\PP$ can be reached at finite time $T$.

{\em Density operator to Density operator Controllable (DDC) in $T$ steps:} if for every initial state $\rho_0\in \DD$ there exist a choice of controls such that   any 
other $\rho_f\in\DD$ can be reached in finite time $T$.

Analogous definition can be given for {\em{Pure state to Density operator Controllable (PDC)}} and 
{\em{Density operator to Pure state Controllable (DPC)}}. Clearly, being $\PP \subset \DD$, it holds that:
\[ DDC \implies DPC \implies PPC,\] \[DDC\implies PDC\implies PPC.\]
\noindent Weaker (approximate) versions of the same controllability properties are of particular interest when dealing with discrete-time systems coming from sampling continuous time models. In fact, 
for these models, there are results correlating the continuos time controllability with the discrete time one, see Section \ref{openloop} below.

It is also possible to think to some notions of {\em{dynamical propagator controllability}}, where, instead of 
looking at the problem of steering a given initial state to a fix final one, we look at the possibility 
 of realizing some given dynamical maps.  We say that a system is:

{\em Unitary controllable (UC) in $T$ steps:} if given any $U\in\UU$ there exist a choice of controls 
  that   realizes  the unitary evolution given by equation (\ref{unitaria})  as a composition 
  of  $T$ evolutions $\Ec_i$, i.e. 
  \[
  U\rho U^{\dag}= \Ec_T\circ\cdots\circ\Ec_1 (\rho), \  \  \   \   \  \forall \rho\in \DD\]
(where $\Ec_i(A)=\Ec(A,\vec{u}_i)$).  

{\em Kraus map controllable (KC) in $T$ steps:} if given any $\Ec$ (see equation (\ref{eq:KrausMap})) 
 exists a choice of $T$ controls  such that $\Ec = \Ec_T\circ\cdots\circ\Ec_1$.

Some immediate relationships between the notions are:
\[ KC \implies DDC,\quad\quad KC \implies UC \implies PPC. \]
The first implication has also been highlighted in \cite{kraus-controllability}. It can be easily derived considering a {\em constant} mapping from $\DD$ to $\rho_f\in\DD,$ $\rho_f$ being the target state. This map can be extended to a linear CPTP map on $\BB$, and hence it admits an OSR (by Kraus-Stinespring theorem \cite{kraus,nielsen-chuang}).

\section{On open-loop discrete-time controllability}\label{openloop}

Controllability results for open-loop, coherent control are well established, and they are essentially based on the Lie-algebra rank condition. 
Moreover, the ability of realizing arbitrary unitary operators in finite time will be key to the results on feedback controllability. This section is devoted to discuss under which conditions this can attained, at least approximately. Consider the controlled Landau-von Neumann equation (\ref{lvn}), 
 the system is then controllable in continuous time if the Jurdievic-Sussman 
 Lie-algebraic rank condition is satisfied \cite{dalessandro-book}.

\begin{thmsec} 
\label{thm:jurdj-susss} The system
\be{lvn1}
 \dot\rho(t)=-i[H_0+\sum_{j=1}^m H_ju_j(t),\rho(t)],
 \ee
 is controllable if and only if the Lie-algebra generated by the Hamiltonians, $ {\rm Lie} \{ -i H_0 , \, -i H_1, \ldots ,-i H_m\},$ is the full $\mathfrak{su}(N)$. 
\end{thmsec}

This condition is generic even with a single control field, that is, almost every pair of drift and control Hamiltonian, $H_0$ and $H_1$, ensures that the associated control Lie algebra is the full $\mathfrak{su} (N)$ \cite{altafini-controllability}. 

Let us introduce the discrete-time model by forcing the control functions $u_j(t)$ to be piece-wise constant on intervals long $\delta$ (in the terminology of \cite{sontag-2}, they are {\em sampled control functions}), and considering the associated evolution:
\be{sampling} \rho(t+\delta)=U(\vec{u}(t))\rho(t)U^\dag(\vec{u}(t)),\ee
where $U(\vec{u}(t))=T{\rm exp}\left(\int_0^\delta H(\vec{u}(t+\tau))d\tau\right),$ and $T{\rm exp}$ denotes the formal, or time-ordered, exponential.
Let us call $\reach_T^\delta(\rho_0)\subseteq\DD$ the set of states reachable from $\rho_0$ by sampled control functions in $T$ steps, and
 \[\reach^\delta(\rho_0)=\bigcup_{T=1}^\infty \reach_T^\delta(\rho_0).\] We say that the system \eqref{lvn} is {\em sampled controllable} (either sampled PPC, PDC, DPC or DDC) if for every pair of states $\rho_0,\rho_f$ (in $\PP$ or $\DD$ according to the type of controllability considered) there is a sample time $\delta$ such that $\rho_f$ is in $\reach^\delta(\rho_0),$ while it is
{\em approximately sampled controllable} if for every pair of states $\rho_0,\rho_f$ there is a sample time $\delta$ such that $\rho_f$ is contained in the closure of $\reach^\delta(\rho_0).$ Sontag proved the following results on the relationship between (continuous-time) controllability and sampled controllability \cite{sontag-1,sontag-2}.

\begin{thmsec}\label{sontag1} If a dynamical system on a simply connected group is controllable (in continuous time), then it is sampled controllable. 
\end{thmsec}

\begin{thmsec} If a dynamical system is controllable (in continuous time), then it is approximately sampled controllable. 
\end{thmsec}

In our setting, considering the dynamical equation \eqref{lvn1}, Theorem \ref{sontag1} ensures that if the system is continuous-time controllable, we can obtain any unitary operator in a finite number of discrete set by sufficiently fast sampled control. An open problem concerns establishing estimates of the time needed to realize a given unitary transformation, and how the sample time may depend on the degree of the accuracy we require for approximate sampled controllability.

\section{Results on Feedback Controllability}\label{sec:feedback}

\subsection{Discrete-time feedback control and background}

We introduce here a discrete-time, Markovian feedback control scheme  \cite{Belavkin_1983_TheoryofControl,viola-engineering,james-robust}, that has been recently studied in depth in \cite{bolognani-arxiv} focusing on stabilization problems. Assume that we can: 
\begin{itemize}\item[(i)] Enact a {\em fixed}, given generalized measurement associated to an OSR $\{ M_k \};$ \item[(ii)] Engineer a set of arbitrary unitary control action $U_k(t)\in\UU$ at each time $t,$ choosing $U_k$ when the $k$-th outcome of the measurement is obtained.
\end{itemize}

Thus, if the state at time $t$ was $\rho(t),$ the state at time $t+1$ conditioned to the $k$-th outcome of the generalized measurement is:
$$\rho(t+1)|_k=\frac{U_k(t)M_k\rho(t)M_k^\dag U_k^\dag(t)}{\tr(M_k^\dag M_k\rho(t))}.$$
Hence, averaging over the possible outcomes we get:

\be{feedback} \rho(t+1)=\sum_kU_k(t)M_k\rho(t)M_k^\dag U_k^\dag(t). \ee

We next recall a characterization of the OSRs that can be realized by exploiting these control capabilities \cite{bolognani-arxiv}. This and the following results heavily rely on a canonical form of the QR decomposition that is recalled in Appendix \ref{canonicalform}.
\begin{prosec}\label{simul}
A measurement with associated operators $\{N_k\}_{k=1}^m$ can be simulated by a certain choice of unitary controls from a measurement $\{M_k\}_{k=1}^m,$ if and only if there exist a reordering $j(k)$ of the first $m$ integers such that:
$$
	{\mathcal F}(N_k)={\mathcal F}(M_{j(k)}),
$$
where ${\mathcal F}$ returns the canonical $R$ factor of the argument, as described in Appendix \ref{canonicalform}.
\end{prosec}


The potential of the feedback strategy for pure state preparation is established by the following \cite{bolognani-cyprus,bolognani-arxiv}.

\begin{thmsec}\label{th:stabilization}
Consider a subspace orthogonal decomposition $\Hi_I=\Hi_S\oplus\Hi_R,$ $\dim(\Hi_S)=1$, and a given generalized measurement associated to Kraus operators $\{M_k\}.$ Let $\{R_k\}$ be the canonical $R$-factors associated to $\{M_k\}$ in a basis consistent with the Hilbert space decomposition above.
The task of achieving global asymptotic stability of $\rho_S=\Pi_S$ by a feedback unitary control policy is feasible if and only if there exists a $\bar k$ such that:
\beq[\rho_S,R_{\bar k}]\neq0.\label{noncommuting}\eeq
\end{thmsec}
Notice that if a pure state is globally asymptotically stabilizable, it means that it belongs to the closure of $\reach^\delta (\rho_0)$ for any initial state $\rho_0 \in \DD$. In the next sections we will use this fact to link the feedback stabilization problem to feedback controllability problems.

\subsection{Three examples}

We here present three examples that will provide motivation for the study of feedback controllability, and counterexamples to generic, finite-time DPC (and hence DDC) and KC properties. Yet, they will suggest some natural questions about weaker controllability properties. Let us agree that $\reach(\rho)$ denotes the reachable set from $\rho$.

{\em Example 1: Feedback-controllability is different from open-loop controllability.}
 An extreme example is the following: Consider a completely depolarizing channel ${\cal E}$ for a two-level system, with $\{M_0=\smallfrac{1}{\sqrt{4}}I,M_1=\smallfrac{1}{\sqrt{4}}\sigma_x,M_2=\smallfrac{1}{\sqrt{4}}\sigma_y,M_3=\smallfrac{1}{\sqrt{3}}\sigma_z\}.$ After a single application of the measurement, the {\em average state} is projected onto $\rho=\frac{1}{2}I,$ the completely mixed state. No subsequent, {\em unconditional} choice of control $U$ has any effect on the dynamics: for $t\geq 0,$
\[\rho(t+1)=U{\cal E}(\rho(t))U^\dag = U\left(\sum_kM_k\rho(t)M_k^\dag\right)U^\dag=\frac{1}{2}I.\] This means that $\reach(\rho_0)=\{\frac{1}{2}I\},$ for every $\rho_0$.
On the other hand, if {\em conditional} controls $\{U_k\}$ are allowed, it is easy to see that choosing e.g. $\{U_k=\sqrt{4}\bar{U} M_k^\dag\},$ we get
\[ \rho(t+1)=\sum_kU_kM_k\rho(t)M_k^\dag U_k^\dag=\bar{U}\rho(t)\bar{U}^\dag.\]
Hence, at least the set of $\rho$ {\em isospectral} to the initial condition $\rho(0)$ is in the reachable set:
\[\{\rho\in\DD|\rho=U\rho(0)U^\dag,\,U\in\UU\}\subset\reach(\rho(0)).\]
However, even the feedback control strategy we are considering has its limitations. A key one is the time needed to reach the desired state, in particular pure states.

{\em Example 2: Feedback purification cannot in general be obtained in finite time.} Consider a full rank state $\rho(t)>0.$ Assume that the generalized measurement we consider has OSR $\{M_1,M_2\},$ with at least $M_1$ is full rank. Then for any control choice $\{U_{1,2}\},$ we have that:
\[U_1M_1\rho(t)M_1^\dag U_1^\dag>0,\]
while $U_2M_2\rho(t)M_2^\dag U_2^\dag\geq 0.$ Hence, being a sum of a strictly positive operator and a positive semidefinte one, $\rho(t+1)>0.$ By iterating the above reasoning, we get that $\rho(t)>0$ for any $0\leq t <\infty.$ Thus, no state on the border of $\DD$ can be reached in finite time from a generic state. The following generalization of this example is in fact immediate:
\begin{prosec}
Consider a feedback controlled system as in \eqref{feedback}. If at least one of the $M_k$s in the OSR has full rank, then no state on $\delta\DD$ is reachable in finite time from $\DD\setminus \delta\DD.$
\end{prosec}

One is then lead to ask: {is the controlled system at least asymptotically DPC? Is there a set of conditions under which the system can be rendered DPC in finite time?} We will prove that feedback discrete-time quantum dynamics are generically {\em asymptotically} (or approximately, in the definition given in Section \ref{openloop}) controllable. In Section \ref{k2} we provide some conditions on the measurement OSR that ensure that feedback system is both {\em DPC and PDC} in finite time.

{\em Example 3: Feedback control does not ensure Kraus-map controllability.}
Consider two CPTP maps on a two-level system, with OSRs
\[M_1=\sqrt{p}\left[\begin{array}{cc} 1 & 0 \\ 0 & -1 \end{array}\right],
M_2=\sqrt{1-p}\left[\begin{array}{cc} 1 & 0 \\ 0 & 1 \end{array}\right],\]
and
\[N_1=\left[\begin{array}{cc} 0 & a \\ 0 & 0 \end{array}\right],
N_2=\left[\begin{array}{cc} 1 & 0 \\ 0 & \sqrt{1-a^2} \end{array}\right],\]
with $1\geq a>0.$  Note that the first OSRs elements are scalar multiples of unitaries, and hence they both have scalar matrices as canonical $R$-factors, while the second OSR is already in canonical form. Assume we want generate the CPTP map associated with $\{N_1,N_2\}$ by feedback control as in \eqref{feedback}.
The canonical $R$-factors being different, the only hope is to feedback-enact an OSR that is equivalent to $\{N_1,N_2\}.$
However, it is immediate to see that for any $U_1,U_2\in\UU$ the dynamical map
\[\rho(t+1)=\sum_{k=1,2}U_kM_k\rho(t)M_k^\dag U_k^\dag\]
remains unital, while the one associated to $\{N_1,N_2\}$ is not.

At a first look, this may seem in contrast with previous results: for example, the main result in \cite{viola-engineering} shows how to feedback engineer arbitrary measurements on the system of interest by using an ingenuous combination of ancillary systems, simple interaction Hamiltonians, projective measurements and fast-pulse control. The attained result is a {\em weaker} KC property, that
needs more general control capabilities, including (essentially) the ability of changing the measurement action, and ensures that the enacted dynamics corresponds in general to the desired one {only at lower-order (in time)}. {\em  What is the class of CPTP maps one can realize via feedback?} A partial answer, of course, is given by Proposition \ref{simul}. However, due to the non uniqueness of the OSR, the fact that the target CPTP map has an OSR that in canonical form is the same of the measurement used in the feedback loop is only sufficient for its realizability by means of a feedback protocol. Providing conditions for exact KC, or characterizing the reachable set of propagators are, to the best of our knowledge, open problems.

\subsection{Generic asymptotic controllability}

Enforcing generalized measurements on the system, one does not lose pure state controllability.

\begin{lemsec}\label{PPC1} Assume that the controlled system dynamics is described by \eqref{feedback}. Then the system is PPC in one step.
\end{lemsec}
\bpr Consider a pure initial state $\rho=\ket{\psi}\bra{\psi}$ and a target $\rho_f=\ket{\phi_f}\bra{\phi_f}.$ The state conditioned on the $k$-th outcome of the measurement step is then $\rho|_k=\ket{\phi_k}\bra{\phi_k},$ with $\ket{\phi_k}=M_k\ket{\psi}/\sqrt{\bra{\psi}M_k^\dag M_k \ket{\psi}}.$ Then to reach $\rho_f$ is sufficient to consider a set of control actions $\{U_k\}$ such that $U_k\ket{\phi_k}=\ket{\psi_f}$  for each $k$.
\epr

Can we always prepare a given pure state starting from an arbitrary density matrix? The answer is generically positive, at least asymptotically, if we allow for feedback control.

\begin{thmsec} \label{purif} Assume the system dynamics to be described by \eqref{feedback}, with a fixed measurement with OSR $\{ M_k \}$ and arbitrary conditional control actions $\{U_k\}\subset\UU$. Then the system is approximately DPC if and only if there is a $k$ such $M_k\neq q V_k,$ for every $q\in\C.,$ and $V_k\in\UU.$
\end{thmsec}
\bpr
As a first step, by properly constructing a basis and invoking  Theorem \ref{th:stabilization}, we will first show that a pure state is stabilizable if $M_k\neq q V_k.$ This condition implies that the corresponding canonical $R$-factor is $R_k\neq q I.$ Let us consider two cases:\\
A) If at least one of the canonical factor is not diagonal, i.e. there exists an element $r_{j\ell}\neq 0$ with $j<\ell,$ reorder the basis so that the $j$-th basis vector becomes the first, and the $\ell$-th is the second. Since the two corresponding columns in $R_k$ were not orthogonal, they will remain so after the change of basis. Hence, when computing {\em again} the canonical $R$-factor, the upper-right $2\times2$ block will be in the form
$$R_k=\left[\begin{array}{cc|c} a & b & * \\0 & d & *\\ \hline 0 & 0 & *\end{array}\right],$$
with $b\neq0.$ According to Theorem \ref{th:stabilization}, the state $\rho_f =\ket{\psi}\bra{\psi},$ $\ket{\psi}=(1,0,\ldots,0)^T$ in the new basis, can be made globally asymptotically stable.\\
B) If all the $R_k$s are diagonal, but at least one is not a scalar matrix, we can find a reordering of the basis so that the upper-right $2\times2$ block of the one $R_k$ is in the form
$$R_k=\left[\begin{array}{cc|c} a & 0 & * \\0 & d & *\\ \hline 0 & 0 & *\end{array}\right],$$
with $a\neq d$. Let us consider a further unitary change of basis $V$ (acting on the right of $R_k$, that modifies the upper left block of $R_k$:
$$V=\frac{1}{\sqrt{2}}\left[\begin{array}{cc|c}1 & 1 & 0 \\1 & -1 & 0  \\\hline 0 & 0 & I\end{array}\right],\;R_kV=\frac{1}{\sqrt{2}}\left[\begin{array}{cc|c} a & a & * \\d & -d & * \\\hline 0 & 0 & * \end{array}\right].
$$ 
Now, being $a\neq d$,its first two columns become not orthogonal. If we compute the canonical R-factor $R'_k$ of $R_k V,$ its first two columns become not orthogonal, and hence $$R'_k=\left[\begin{array}{cc|c} a' & b' & 0 \\0 & c' & 0 \\\hline 0 & 0 & * \end{array}\right],
$$ 
with $b'\neq 0.$ Notice that the construction above works also for $d=0.$ Thus the state $\rho'_f=\ket{\psi'}\bra{\psi'},$ $\ket{\psi'}=(1,0,\ldots,0)^T$ in the new basis can be made asymptotically stable by feedback control. \\
To conclude the ``if'' implication, assume we reach a $\varepsilon$-neighborhood (in trace distance) of $\rho'_f$ at some time $T-1$. Then on the $T$ step we can apply a different set of unitary control actions, as in Lemma \ref{PPC1}, which realize the one-step transition $\rho'_f$ to $\rho_f,$ and since CPTP maps are trace norm contractions we end up in a $\varepsilon'$-neighborhood (in trace distance) of $\rho_f,$ with $\varepsilon'\leq\varepsilon$. 
On the other hand, assume that $R_k=q_kI,\;\forall k.$ Then the feedback dynamics \eqref{feedback} becomes:
\[\rho(t+1)=\sum_k q_k^2 U_k\rho(t)U_k^\dag.\]
A map of this form can only reach states in the convex hull of the set isospectral to $\rho(t).$  Hence if $\rho(0)$ is in the interior of $\DD,$ the closure of the reachable set $\reach(\rho(0))$ cannot contain any pure state.
\epr
It is worth remarking that: (i) the proof is constructive, since it implicitly uses the constructive result of \cite{bolognani-arxiv}; (ii) relying on a stabilization procedure, the control strategy is {\em robust} with respect to uncertainty on the initial state $\rho_0$; (iii) the time needed for approximately reaching an $\varepsilon$-neighborhood of the target state can be estimated by computing the slowest eigenvalue of the feedback-controlled map. (iv) the condition $M_k\neq qV_k$ for some $k$ is generic, and it fails only for probabilistic average of unitary effects. In other words, the class of {\em measurements} that do not allow for DPC are those  that are associated to an average over the conditonial states of the form:
\beq\label{unitmeas}\hat\rho=\sum_k p_k U_k\rho U_k^\dag.\eeq
Furthermore, this corollary of Proposition \ref{purif} comes at no cost:
\begin{corsec} Assume that we can control the system as in Theorem \ref{purif} above. Then asymptotic feedback purification of the state can be achieved if and only if there is a $k$ such $M_k\neq q U_k,$ for every $q\in\C.,$ and $U_k\in\UU.$
\end{corsec}

\noindent The results above in turn imply that feedback makes the system DDC, provided that we can randomly choose the unitary controls in a finite set with given probabilities:

\begin{corsec}  Assume that we can control the system as in Theorem \ref{purif} above, and in addition we can pick a control action at random from a finite set $\{\hat U_j\}$ with an {\em arbitrary} probability distribution $\{p_j\}$.  Then the system is approximately DDC if and only if there is a $k$ such $M_k\neq q V_k,$ with $q\in\C.,$ and $V_k\in\UU.$\end{corsec}

\bpr By Theorem \ref{purif}, there exists a finite time $T$ so that we can get arbitrarily close to a pure state $\ket{\psi}\bra{\psi}.$ Assume the target state is $\rho_f=\sum_j p_j \ket{\phi_j}\bra{\phi_j},$ and define the control actions $\hat U_j$ so that $\hat U_j\ket{\psi}=\ket{\phi_j}.$ Than at some time $T$ it suffices to extract at random a $\hat U_j$ with probability $\{p_j\},$ so that the average dynamics (disregarding which $\hat U_k$ has been extracted, gives 
\[\sum_jp_j\hat U_j\ket{\psi}\bra{\psi}U_j^\dag=\sum_k p_j \ket{\phi_j}\bra{\phi_j}=\rho_f.\] 
\epr
Notice that, up to the last step, the choice of the unitary control actions is time-independent, that is, at each iteration the average dynamics is represented by the same OSR:
\[\Ec(\rho(t)) = \sum_kU_kM_k\rho(t)M_k^\dag U_k^\dag.\]

\section{Sufficient conditions for finite-time state controllability}\label{k2}

Assume that a certain generalized measurement has only two outcomes, and associated operators $M_1$ and $M_2$ such that:
\be{uno}
 M_1^\dag M_1 + M_2^\dag M_2 =I.\ee
Moreover, assume:
\begin{enumerate}
\item {\em Both matrices are diagonal;} 
\item {\em Both matrices are singular.}
\end{enumerate}
Assumption 1) is not restrictive under feedback control assumptions, as it shown in the following lemma. 
\begin{lemsec}Consider two generic $\tilde M_1, \tilde M_2$ that satisfy \eqref{uno}. Then there exist a unitaries $U_0,U_1,U_2$ such that $M_j=U_j\tilde M_jU_0$ is diagonal for $j=1,2.$
\end{lemsec}
\bpr
By appropriately choosing the reference basis through a unitary $U_0,$ and a enacting a (feedback) unitary $U_1,$ we can diagonalize any $\tilde M_1,$ by e.g. singular value decomposition $U_1 \tilde M_1 U_0 = M_1=\rm{diag}(\alpha_1,\ldots,\alpha_N).$ Then $\Delta=U_0^\dag\tilde M_2^\dag \tilde M_2 U_0$ must be diagonal, since \eqref{uno} holds, and hence it admits a diagonal square root of the form $U_2\tilde M_2 U_0.$
\epr
Given assumption 1)-2), without loss of generality, the two matrices $M_i$ have then the following form with respect to a reference basis $\{\ket{e_j}\}_{j=1}^N$:
\be{forma}
M_1 = \left[ \begin{array}{ccccc}
   0 & 0 & 0 & \cdots & 0 \\
  0 & \alpha_2 & 0 & \cdots & 0 \\
  0 & 0 & \alpha_3 & \cdots & 0 \\
  \vdots&\vdots&\vdots&\vdots&\vdots \\
  0 & 0 & 0 & \cdots & \alpha_{N} \end{array} \right],\,
  M_2 = \left[ \begin{array}{ccccc}
  \beta_1 & 0 & 0 & \cdots & 0 \\
  0 & 0 & 0 & \cdots & 0 \\
  0 & 0 &\beta_3 & \cdots & 0 \\
  \vdots&\vdots&\vdots&\vdots&\vdots \\
  0 & 0 & 0 & \cdots & \beta_{N} \end{array} \right] 
    \ee
    where, to satisfy \eqref{uno}, we must have $|\alpha_2|=|\beta_1|=1$ and, for $i=3,\ldots, N$,  $|\alpha_i|^2+|\beta_i|^2=1$. It is immediate to see that a measurement in this form is able to {\em distinguish with certainty} at least the first two orthogonal states of the basis in which $M_1,M_2$ have the form \eqref{forma}. We can now prove that the feedback controlled dynamics is finite-time DPC.
   
   \bp{prima}
    There exists a choice $U_1(0),\cdots,$ $U_1(N-2)$ and $U_2(0),\cdots, U_2(N-2)$, such that for
   any  $\rho_0=\sum_{i=1}^N \gamma_i |v_i\rangle\langle v_i|$,  $\rho(N)$ is a pure state.
   \ep
   
   \bpr
Let $w$ be the unit vector such that the target state   $\rho(N)$ is equal to $|w\rangle\langle w|$.
For $i=0,\ldots, N-3$,  define the two matrices $U_1(i)$ and $U_2(i)$ as the permutations matrices defined by the following relationships:
\be{controlli1}\begin{array}{ll}
U_1(i)\ket{e_1}=\ket{e_{N-i}} &    U_2(i)\ket{e_2}=\ket{e_{N-i}}   \\
U_1(i)\ket{e_{N-i}}=\ket{e_{1}} &     U_2(i)\ket{e_{N-i}}=\ket{e_{2} }  \\
U_1(i)\ket{e_j}=\ket{e_{j}},\;   j\neq 1, N-i,&\\ U_2(i)\ket{e_j}=\ket{e_{j}}, \; j\neq 2, N-i &\end{array}
\ee
and let $U_1(N-2)$ and $U_2(N-2)$ be any two unitary matrices such that  
\be{controlli2}
U_1(N-2)\ket{e_2}=\ket{w}, \ \ \  \   \  U_2(N-2)\ket{e_1}=\ket{w}.
\ee
We first prove by induction on $k=0,\ldots,N-2$, that $\rho(k)$ is of the following type:
\be{tipo}
\rho(k)=\sum_{j=1}^{l_k}b^k_j  |z^k_j\rangle\langle z^k_j|,  \ \text{ with } \
\ket{z^k_j}=\sum_{s=1}^{N-k} (c^k_j)_s \ket{e_s},
\ee
that is, at step $k$ the state $\rho_k$ has support only on the subspace generated by the first $N-k$ basis vectors, $\{\ket{e_j}\}_{j=1}^{N-k}.$
For $k=0$ the statement is trivial, so assume that (\ref{tipo}) holds for $k< N-2$, then
\[
\rho(k+1)=\sum_{i=1}^2 U_i(k)M_i\rho(k)M_i^\dag U_i(k)^\dag .\]
We have, for $i=1,2$ :
\[
U_i(k)M_i\rho(k)M_i^\dag U_i(k)^\dag{\hspace{-2mm}}={\hspace{-2mm}} \sum_{j=1}^{l_k}b^k_j  U_i(k)M_i|z^k_j\rangle\langle z^k_j|M_i^\dag U_i(k)^\dag
\]
Moreover it holds:
\[
U_1(k)M_1|z^k_j\rangle= \sum_{s=1}^{N-k} (c^k_j)_s U_1(k)M_1|e_s\rangle =\]
\[ \sum_{s=2}^{N-k} (c^k_j)_s \alpha_s  U_1(k)|e_s\rangle
 = \hspace{-2mm}
\sum_{s=2}^{N-k-1} (c^k_j)_s \alpha_s  |e_s\rangle +
\]
\[+ (c^{k}_j)_{N-k} \alpha_{N-k}|e_1\rangle 
 =
\sum_{s=1}^{N-k-1} ( \hat{c}_{j}^{k+1})_s|e_s\rangle= \ket{\hat{z}^{k+1}_j}.
\]
Using the same argument and exchanging $1$ with $2$, we get:
\[
U_2(k)M_2|z^k_j\rangle= 
\sum_{s=1}^{N-k-1} ( \tilde{c}_{j}^{k+1})_s|e_s\rangle= \ket{\tilde{z}^{k+1}_j}.
\]
Thus equation (\ref{tipo}) holds  for $k+1$ with:  
\[
l_{k+1}=2l_k  \ \text { and } \left\{ \begin{array}{ll}
                        z^{k+1}_j = \hat{z}^{k+1}_j   & \text{for } j=1,\ldots,l_k, \\
                         & \\
                        
                         z^{k+1}_j = \tilde{z}^{k+1}_{j-l_k} & \text{for } j=l_k,\ldots,l_{k+1}. \\ 
                         \end{array} \right.
                         \]
Using (\ref{tipo}) for $k=N-2$, and letting  omitting for simplicity  the index $N-2$, we have:
\[
\rho(N-2)=\sum_{j=1}^{l }\beta_j  |z_j\rangle\langle z_j|,
\]
  with 
$\ket{z_j}=(c_j)_1\ket{e_1}+(c_j)_2 \ket{e_2}$.
Now we have:
\[
\rho(N-1) =            
\sum_{i=1}^2\sum_{j=1}^{l }\beta_j U(N-2)M_i |z_j\rangle\langle z_j|M_i^\dag U(N-2)^\dag.
\]
Moreover:
\[\begin{split}
U(N-2)M_1 |z_j\rangle&=U(N-2)M_1\left((c_j)_1|e_1\rangle+(c_j)_2|e_2\rangle\right)\\&=
U(N-2)(c_j)_2|e_2\rangle= (c_j)_2|w\rangle,\end{split}
\]
and, analogously, 
\[\begin{split}
U(N-2)M_2 |z_j\rangle&=U(N-2)M_2\left((c_j)_1|e_1\rangle+(c_j)_2|e_2\rangle\right)\\&=
U(N-2)(c_j)_1|e_1\rangle= (c_j)_1|w\rangle.
\end{split}\]
Thus, summing up, we obtain:
\[
\rho(N_1) =            
\sum_{j=1}^{l }\beta_j \left( |(c_j)_2|^2 +|(c_j)_1|^2 \right) |w\rangle \langle w| =
 |w\rangle \langle w|,
\]
as desired.
   \epr
   
   The converse is also true, that is, the system is finite-time PDC.
   
   \bp{prima1}
    Assume that $\rho_0$ is a pure state, then for any 
    $\rho_f= \sum_{i=1}^N \gamma_i |v_i\rangle\langle v_i|$,  there exists a sequence of controls of length $N$ that steers $\rho_0$ to $\rho_f$.   \ep
    
    \bpr The explicit construction of a set of effective controls can be done  following the procedure detailed below.\\
    {\em{First step:} Prepare an appropriate pure state.}     
 Assume that $\rho_0= |w\rangle \langle w|$.  If $M_i |w\rangle=0$, then let $U_i(0)$  be any unitary matrix,  if $M_i |w\rangle\neq 0$, then let 
  $U_i(0)$ be any unitary matrix  such that 
 \[
 U_i(0)\left( \frac{ M_i |w\rangle }{ ||M_i |w\rangle || } \right)  =
 \sqrt \gamma_1 |e_2\rangle +\sqrt{(1-\gamma_1)}|e_1\rangle =|z_1\rangle.
 \]
Then, $\rho(1)$ is again a pure state and  we have 
$\rho(1) =  |z_1\rangle\langle z_1|.$\\
{\em{Second Step}: Preparing the first element.} 
Let $U_1(1)$ be any unitary matrix such that $U_1(1)|e_2\rangle = |e_N\rangle$, and 
$U_2(1)$ be any unitary matrix such that 
\[
U_2(1)|e_2\rangle = \sqrt {\frac{\gamma_2}{1-\gamma_1}} |e_2\rangle + \sqrt  {1-\frac{\gamma_2}{1-\gamma_1}} |e_1\rangle =|z_2\rangle.
\]
\noindent With  this choice we have:
\[ 
\rho(2)= \sum_{i=1}^2U_i(1)M_iz_1\rangle\langle z_1| M_i^\dag U_i(1)^\dag = \]
\[
=\gamma_1 U_1(1)|e_2\rangle\langle e_2|U_1(1)^\dag +
(1-\gamma_1)U_2(1)|e_1\rangle\langle e_1|U_2(1)^\dag\]
\[ =
 \gamma_1 |e_N\rangle\langle e_N | +
(1-\gamma_1)|z_2\rangle\langle z_2|.
\]
{\em{Successive steps}.}
Notice that $e_N$ and $z_2$ are orthogonal. Let  
$U_1(2)$ be such that:
\[
U_1(2) |e_N\rangle=|e_N\rangle, \ U_1(2) |e_2\rangle=|e_{N-1}\rangle, 
\]
and let $U_2(2)$ be such that:
$
U_2(2) |e_N\rangle=|e_N\rangle$,  and
\[
 U_2(2) |e_1\rangle =  \sqrt {\frac{\gamma_3}{1-(\gamma_1+\gamma_2)}} |e_2\rangle +
 \]
 \[ \sqrt  {1-\frac{\gamma_3}{1-(\gamma_1+\gamma_2)}} |e_1\rangle =|z_3\rangle. 
\]
Now we have:
\[
\rho(3)=
\]
\[ U_1(2)M_1 \left(   \gamma_1 |e_N\rangle\langle e_N | +
(1-\gamma_1)|z_2\rangle\langle z_2| \right) M_1^\dag U_1(2)^\dag +
\]
\[
U_2(2)M_2 \left(   \gamma_1 |e_N\rangle\langle e_N | +
(1-\gamma_1)|z_2\rangle\langle z_2| \right) M_2^\dag U_2(2)^\dag
 \]
 \[
= \gamma_1|\alpha_N|^2|e_N\rangle\langle e_N| + \gamma_2 |e_{N-1}\rangle\langle e_{N-1} |
 + \gamma_1|\beta_N|^2|e_N\rangle\langle e_N| \]\[ +(1-(\gamma_1+\gamma_2))|z_3\rangle\langle z_3|=
\]
\[
=  
 \gamma_1 |e_N\rangle\langle e_N | +  \gamma_2 |e_{N-1}\rangle\langle e_{N-1} | +
(1-(\gamma_1+\gamma_2))|z_3\rangle\langle z_3|.
\]

Iterating this construction, after $N-1$ steps, we will get:
\[
\rho(N-1)=  \gamma_1 |e_N\rangle\langle e_N | +  \gamma_2 |e_{N-1}\rangle\langle e_{N-1} | +
\ldots +
\]
\[ +\gamma_{N-2} |e_3\rangle\langle e_3| 
  + \left(1-\left(\sum_{i=1}^{N-2}{\gamma_i}\right)\right) |z_{N-1}\rangle \langle z_{N-1}|,
\]
with
\[\begin{split}
|z_{N-1}\rangle&= 
\sqrt {\frac{\gamma_{N-1}}{1-\left(1-\left(\sum_{i=1}^{N-2}{\gamma_i}\right)\right)}} |e_2\rangle \\&+ \sqrt  {1-\frac{\gamma_{N-1}}{1-\left(1-\left(\sum_{i=1}^{N-2}{\gamma_i}\right)\right)}} |e_1\rangle 
\end{split}\]
{\em{Final two steps:} finalizing the construction.} 
Now, letting $U_1(N-1)=U_2(N-1)=I$, we obtain:
\[
\rho(N) = \gamma_1 |e_N\rangle\langle e_N | +  \gamma_2 |e_{N-1}\rangle\langle e_{N-1} |\]
\[ +
\ldots \gamma_{N-1} |e_2\rangle\langle e_2| +
  \left(1-\left(\sum_{i=1}^{N-1}{\gamma_i}\right)\right) 
|e_1\rangle \langle e_1|= 
\]
\[
=\sum_{i=1}^N \gamma_i |e_{N-i+1}\rangle\langle e_{N-i+1} |.
\]
Next, define $U_1(N)=U_2(N)=U$, where $U$ is the unitary matrix such that $U|e_i\rangle=|v_{N-i+1} \rangle$. We then get:
\[
\rho(N+1) =\sum_{i=1}^N \gamma_i |v_i\rangle\langle v_i|,
\]
as desired.\epr

\section{Conclusions and Outlook}

We investigated the controllability properties of discrete-time Markovian quantum dynamics, in particular showing that sampled, open-loop dynamics inherited their controllability properties from the underlying continuous-time models, and that the ability of implementing discrete-time feedback generically allows for complete DDC, but only asymptotically. Furthermore, we have shown that the ability of implementing simple, two-outcome measurements with singular operators makes the system DDC in closed loop and in {\em finite time}. That is, by extracting just one bit of classical information at the time, with at least the ability of discerning two pure states with certainty, we are allowed to feedback-enact arbitrary state-to-state transitions in an $N$-level system, and the time needed to reach the desired state is just $2N.$ This result shows {e.g.} how feedback cooling can be obtained in finite time, and not only asymptotically, for discrete-time evolutions. Our results also complement the ones presented in \cite{bolognani-arxiv,bolognani-cyprus}, and the discussion in Section \ref{openloop} support for the strong assumption of unitary open-loop controllability of the discrete-time system. Some open questions remain, and need further investigation: if the measurement OSR is not of the form \eqref{unitmeas}, is the system feedback KC? Can the strategy presented in Section \ref{k2} be extended to OSRs with more than two operators? 

The presented analysis also suggests that similar results may be pursued for continuous-time QDS. The techniques  developed in \cite{ticozzi-QDS,ticozzi-markovian,ticozzi-generic} for QDS engineering by open-loop and output-feedback control, in the spirit of \cite{wiseman-feedback}, could be used to overcome the generic absence of controllability pointed out by \cite{altafini-markovian,altafini-open}. Another interesting and timely development would be to study how the controllability properties change in presence of locality constraints in multipartite, distributed quantum systems. The use of noise engineering for quantum-computation related tasks has recently received considerable attention from the physics community \cite{kraus-entangled,verstraete2009}, but many fundamental control-theoretic questions remain unanswered.
Lastly, in Section \ref{openloop} we considered the effect of sampling on the controls: another  constraint, worth of further investigation, might be introduced by their quantization. When limiting the choice of the control functions to a countable (or finite) set, the controllability issues appear to be strongly connected with the problem of establishing {\em universality of a set of unitary gates}, and the problem of efficiently (in polynomial time with respect to the number of components of the system) generating the desired gates \cite{deutsch-universal,lloyd-universal,nielsen-chuang}. 
\appendix

\section{A canonical QR decomposition}
\label{canonicalform}

In this appendix we recall some technical results about QR decomposition that lead to a canonical form with respect to the left action of the unitary matrix group.

\begin{dfnsec}[QR decomposition \cite{horn-johnson}]
A QR decomposition of a complex-valued square matrix $A$ is a decomposition of $A$ as
$$
   A = QR,
$$
where $Q$ is a unitary matrix and $R$ is an upper triangular matrix.
\end{dfnsec}

The QR decomposition of a given complex-valued square matrix $A$ is not unique.
In the case of non-singular matrix $A$, one can show that the upper triangular factors of any two QR decompositions of $A$ differ only for the phase of their rows. When $A$ is singular, on the other hand, this is not true. However, introducing some conditions on the R matrix, it is possible to obtain a \emph{canonical form} for the QR decomposition in a sense that will be explained later in this section. The following theorem characterize the canonical QR decomposition and guarantees its existence.
\begin{thmsec} \label{th:QRconstruction}
Given any (complex) square matrix $A$ of dimension $n,$ it is possible to derive a QR decomposition $A= QR$ such that
\begin{equation} \label{eq:characterizationQR}
r_{ij} = 0 \quad \forall j \le n, \forall i>\rho_j
\end{equation}
where $\rho_j$ is the rank of the first $j$ columns of $A$, and such that the first nonzero element of each row of $R$ is real and positive.
\end{thmsec}
The proof of this theorem \cite{bolognani-arxiv} also provides a method to construct such a decomposition, by a variation of the standard orthonormalization approach. Moreover, we can prove that the $R$ obtained in this way is a canonical form. 
Let $\mathcal{G}$ be a group acting on $\C^{n\times n}.$ Let $A,B\in \C^{n\times n}.$ If there exists a $g\in{\cal G}$ such that $g(A)=B,$ we say that $A$ and $B$ are ${\cal G}$-equivalent, and we write $A\sim _{\cal G} B.$ 
\begin{dfnsec}\label{canonicalform1} A canonical form  with respect to $\mathcal{G}$ is a function $\mathcal{F}:\C^{n\times n}\rightarrow \C^{n\times n}$ such that for every $A,B\in\C^{n\times n}$:\begin{itemize}
\item[i.]${\cal F}(A)\sim _{\cal G} A $;
\item[ii.] ${\cal F}(A)={\cal F}(B)$ if and only if $A\sim _{\cal G} B.$
\end{itemize}
\end{dfnsec}
Let us consider the unitary matrix group $\mathcal{U}(n)\subset\C^{n\times n}$ and consider its action on $\C^{n\times n}$ through left-multiplication, that is, for any $U\in\mathcal{U}(n),\,M\in\C^{n\times n}$:
$$U(M)=UM.$$ The following result has been proven in \cite{bolognani-arxiv}.
\begin{thmsec} \label{th:Canonical}
	Define $\mathcal{F}(A)=R$, with  $R$ the upper-triangular matrices obtained by the procedure described in the proof of Theorem \ref{th:QRconstruction}. Then $\mathcal{F}$ is a canonical form with respect to ${\cal U}(n)$ (and its action on $\mathbb{C}^{n\times n}$ by left multiplication).
\end{thmsec}

\section*{Acknowledgements}                               
F.T. gratefully acknowledges Lorenza Viola for stimulating discussions on the topics of this paper. Work partially supported by the CPDA080209/08 and QFuture research grants of the University of Padova, and by the Department of Information Engineering research project ``QUINTET''.

\bibliographystyle{plain}
\bibliography{bibliography}

\end{document}